\documentclass[aps,prl,twocolumn,showpacs,superscriptaddress,a4paper,floatfix]{revtex4-1}
\usepackage[latin1]{inputenc}
\usepackage{graphicx}
\usepackage{amsmath,amssymb}
\def\lesssim{\ \raise.3ex\hbox{$<$}\kern-0.8em\lower.7ex\hbox{$\sim$}\ }
\def\gesim{\ \raise.3ex\hbox{$>$}\kern-0.8em\lower.7ex\hbox{$\sim$}\ }

\newcommand \beq{\begin{eqnarray}}
\newcommand \eeq{\end{eqnarray}}
\begin{document}
\title{Thermal crossover, transition, and coexistence in Fermi polaronic spectroscopies}
\author{Hiroyuki Tajima}
\affiliation{Quantum Hadron Physics Laboratory, RIKEN Nishina Center (RNC),
  Wako, Saitama 351-0198, Japan}
\author{Shun Uchino}
\affiliation{Waseda Institute for Advanced Study, Waseda University, Shinjuku, Tokyo, 169-8050, Japan}
\begin{abstract}
We investigate thermal evolution of radio-frequency (RF) spectra of a spin-imbalanced Fermi gas near a Feshbach resonance
in which degenerate Fermi-polaron and classical Boltzmann-gas regimes emerge in the low-temperature and high-temperature limits, respectively.
By using self-consistent frameworks of strong-coupling diagrammatic approaches,
both of the ejection and reserve RF spectra available in cold-atom experiments are analyzed.
We find a variety of transfers from Fermi polarons to the Boltzmann gas such that
a thermal crossover expected in the weak-coupling regime is shifted to a sharp transition near unitarity
and to double-peak coexistence of attractive and repulsive branches in the strong-coupling regime.
Our theory provides semiquantitative descriptions for a recent experiment on the ejection RF spectroscopy at unitarity~[Z. Yan {\it et al}., arXiv:1811.00481v1] 
and suggests the importance of beyond-two-body correlations in the high-temperature regime due to the absence of Pauli-blocking effects.
\end{abstract}
\maketitle
A spectroscopic method is one of central themes in physics including
hadron-mass spectroscopy  in nuclear physics~\cite{PRD60.091503,PPNP46.459,RMP82.1095} and
gravitational wave detection in astrophysics~\cite{GW170817,AbbottAstrophys}. 
In condensed matter,
a spectroscopic technique is of importance to examine low-energy excitations in
quantum many-body systems, which led to discoveries of pseudogap in high-$T_c$ superconductors~\cite{PRL80.149,RMP75.473,RMP79.353}
and of topological states of matter~\cite{Lu2012}.
In an ultracold atomic gas that is an ideal platform to realize nontrivial quantum states of matter and yet has limited probes due to electrical charge neutrality,  
a quantum many-body spectroscopy is an inevitable tool to extract fundamental properties of the system~\cite{torma}.
For instance, the Bragg spectroscopy allows to measure
Nambu-Goldstone modes in superfluid gases~\cite{steinhauer,Hoinka}, and the lattice modulation spectroscopy to measure 
the Mott gap in an optical lattice system~\cite{soferle}.
In addition, the radio-frequency (RF) spectroscopy in cold atoms provides an alternative route to probe interacting atomic gases, and revealed the essential properties such as  
pseudogap in normal Fermi gases~\cite{Chin,Stewart,Gaebler,Sagi}, superfluid gap~\cite{Schirotzek} and Higgs mode~\cite{Behrle} in superfluid Fermi gases,
and Efimov effect in three-component Fermi gases~\cite{Lompe,Nakajima}. 
Since the RF spectroscopy is sensitive to excitation properties in quantum many-body systems, 
of current interest in the RF spectrum measurements is a polaron which is a prototype on how a strong interaction affects quasiparticle properties. 
\par
In cold atoms, polaron physics can be achieved simply by considering a polarized mixture.
When such a mixture consists of a two-component Fermi gas, the system at a low-temperature reduces to
a Fermi polaron, which is a mobile impurity surrounded by the Fermi sea.
The RF spectral results as well as the experimental realizations~\cite{Schirotzek0902,nascimbene,Kohstall1112,koschorreck2012,Cetina96,PhysRevLett.118.083602}
triggered a number of theoretical works mostly under the condition of a single impurity at absolute zero~\cite{Combescot2008,Bruum2010,PhysRevA.85.021602,PhysRevA.85.033631,PhysRevA.74.063628,Combescot2007,Punk2009,PhysRevA.80.033607,PhysRevA.81.041602,PhysRevLett.106.166404,Trefzger2012,Prokofev20081,Prokofev20082,Vlietinck2013,Kroiss2015,PhysRevA.94.051605,Schmidt:2011zu,PhysRevA.95.013612,Bulgac,Goulko2,massignan}.
The current consensus is that  such single-impurity calculations agree well with
low-temperature spectral data of the experiments at typical impurity concentrations.
\par
Quite recently, the MIT group tracked thermal evolution of the RF spectra of a spin-imbalanced unitary Fermi gas
in a homogeneous box potential~\cite{boxpot} from the low-temperature quantum regime to the  classical Boltzmann-gas regime~\cite{Boiling}.
The observed temperature dependence of the RF spectra indicates the existence of a nontrivial sharp transition between quantum and classical regimes, in addition to a precise determination of the polaron energy.
Obviously, an explanation of the MIT experiment demands self-consistent theoretical frameworks to
take the strong correlations into account~\cite{Massignan2,Massignan3,Pietila,Doggen,Hu,Tajima_NJP,Edward,Mistakidis,Mulkerin}. 
\par
In this Letter, motivated by the MIT experiment~\cite{Boiling}, 
we investigate thermal evolution of RF spectra  of a strongly-interacting polarized Fermi gas 
by means of self-consistent many-body calculations.
By analyzing typical schemes of the RF spectroscopy, 
we demonstrate that a transfer from Fermi polarons to Boltzmann gas becomes rich due to correlation effects and
thermal broadening.
In particular, we find three types of the transfers such as crossover, transition, and  coexistence as a function of the interaction strength.
A comparison with the experiment at unitarity shows that our calculations perfectly reproduce the spectral properties in a low-temperature regime and yet have some deviations in a high-temperature regime, implying the importance of beyond-two-body scattering  processes  due to 
the absence of Pauli-blocking effects of majority fermions.
\par
We consider a spin-imbalanced Fermi gas with a large scattering length $a$ such that
$|k_{\rm F}a|>1$, where $k_{\rm F}$ is  the Fermi momentum.
(in what follows, we use $\hbar=k_{\rm B}=1$ and the system volume is taken to be unity).
In the RF spectroscopy, an RF field is applied to an atomic cloud, which introduces a transition between one of spin states and
another atomic excited state, and the current of the excited state induced by the transition is measured~\cite{torma}. 
Since a position of a Feshbach resonance is different between different internal states,
one can realize a couple of schemes of the RF spectroscopy.
In the ejection RF spectroscopy, a strongly-interacting atomic gas is initially prepared and is transferred to a weakly-interacting final state.
By neglecting the final-state interaction effect and assuming a small Rabi frequency $\Omega_{\rm R}$ to justify the linear response treatment, 
the RF spectum as functions of the energy $\omega$  is given by~\cite{Boiling,torma}
\beq
\label{eq1}
I_{\rm E}(\omega)=2\pi\Omega_{\rm R}^2\sum_{\mathbf{p}}f(\xi_{\mathbf{p},\downarrow}-\omega)A_{\mathbf{p},\downarrow}(\xi_{\mathbf{p},\downarrow}-\omega).
\eeq
Here, $f(\xi)=\left(e^{\xi/T}+1\right)^{-1}$ is the Fermi distribution function, $A_{\mathbf{p},\sigma}(\omega)$ the spectral function, $T$ the temperature, and
$\xi_{\mathbf{p},\sigma}=\frac{p^2}{2m}-\mu_{\sigma}$  the kinetic energy of a Fermi atom
measured from the chemical potential $\mu_{\sigma}$ with momentum $\mathbf{p}$, mass $m$, and pseudospin $\sigma=\uparrow, \downarrow$,
where $\uparrow (\downarrow)$ corresponds to the majority (minority) component. 
In the reverse RF spectroscopy, in contrast, a weakly-interacting atomic gas is initially prepared and is transferred to a strongly-interacting final state.
 In this case, the spectrum in absence of the initial-state interaction is  given by~\cite{PhysRevLett.118.083602,torma}
\beq
\label{eq2}
I_{\rm R}(\omega)=2\pi\Omega_{\rm R}^2\sum_{\mathbf{p}}f(\xi_{\mathbf{p},{\rm i}})A_{\mathbf{p},\downarrow}(\xi_{\mathbf{p},\downarrow}+\omega),
\eeq 
where $\sigma={\rm i}$ in Eq. (\ref{eq2}) indicates the initial state of reverse RF measurements.
The spectral function $A_{\mathbf{p},\sigma}(\omega)=-\frac{1}{\pi}{\rm Im}G_{\mathbf{p},\sigma}(i\omega_n\rightarrow \omega+i\delta)$ is obtained from the analytic continuation of thermal Green's function $G_{\mathbf{p},\sigma}(i\omega_n)=\left[i\omega_n-\xi_{\mathbf{p},\sigma}-\Sigma_{\mathbf{p},\sigma}(i\omega_n)\right]^{-1}$,
where $\omega_n$ and $\Sigma_{\mathbf{p},\sigma}(i\omega_n)$ are the fermionic Matsubara frequency and the self-energy, respectively \cite{noteC}.
We use a Pad\'{e} approximation with a small number $\delta=10^{-2}\varepsilon_{\rm F}$ for the analytic continuation ($\varepsilon_{\rm F}$ is the Fermi energy of majority atoms) \cite{Tajima_NJP}. 
We note that $\mu_{\sigma}$ is determined by solving an equation for the number density
$n_{\sigma}=T\sum_{\mathbf{p},\omega_n}G_{\mathbf{p},\sigma}(i\omega_n)$.
According to the MIT experiment~\cite{Boiling}, the impurity concentration $x=\frac{n_{\downarrow}}{n_{\uparrow}}$ is fixed as $x=0.1$.
\par
\begin{figure}[t]
\begin{center}
\includegraphics[width=8.5cm]{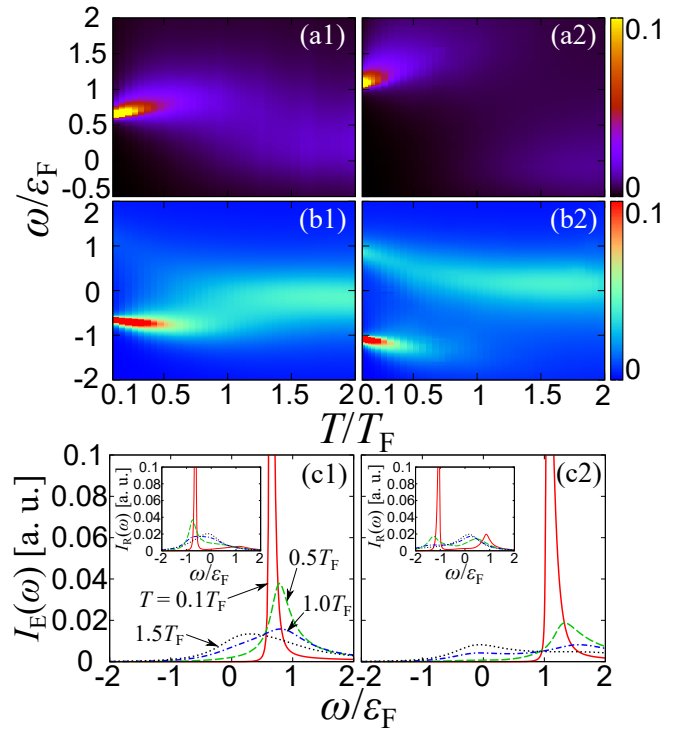}
\end{center}
\caption{Calculated ejection RF spectra $I_{\rm E}(\omega)$ (upper panels) and reverse RF spectra $I_{\rm R}(\omega)$ (lower panels) in arbitrary units for $(k_{\rm F}a)^{-1}=0$ [(a1) and (b1)] and $(k_{\rm F}a)^{-1}=0.4$ [(a2) and (b2)].
$\varepsilon_{\rm F}$ and $T_{\rm F}$ are the Fermi energy and the Fermi temperature of majority atoms, respectively.
(c1) and (c2) show the line-shapes of $I_{\rm E}(\omega)$ [Inset $I_{\rm R}(\omega)$] of $(k_{\rm F}a)^{-1}=0$ and $(k_{\rm F}a)^{-1}=0.4$  at different temperatures.  
}
\label{fig1}
\end{figure}
The many-body correlations associated with the strong interaction are included in the self-energy. In this Letter, we employ the extended $T$-matrix approximation (ETMA) \cite{Tajima_NJP,PhysRevA.86.043622,PhysRevA.89.033617,PhysRevA.93.013610,PhysRevA.96.033614} which can reproduce observed ground-state properties in both imbalanced \cite{Tajima_NJP} and balanced \cite{PhysRevA.95.043625,PhysRevX.7.041004} Fermi gases near the unitarity limit.
In this formalism, $\Sigma_{\mathbf{p},\sigma}(i\omega_n)$ is given by
\beq
\label{eq3}
\Sigma_{\mathbf{p},\sigma}(i\omega_n)=T\sum_{\mathbf{q},\nu_j}\Gamma_{\mathbf{q}}(i\nu_j)G_{\mathbf{q}-\mathbf{p},-\sigma}(i\nu_j-i\omega_n),
\eeq
where 
\beq
\label{eq4}
\Gamma_{\mathbf{q}}(i\nu_j)=\left[\frac{m}{4\pi a}+\Pi_{\mathbf{q}}(i\nu_j)-\sum_{\mathbf{p}}\frac{m}{p^2}\right]^{-1}
\eeq
is the many-body $T$-matrix with the scattering length $a$
and $\nu_j$ is the bosonic Matsubara frequency.
The pair susceptibility $\Pi_{\mathbf{q}}(i\nu_j)$ is given by 
\beq
\label{eq5}
\Pi_{\mathbf{q}}(i\nu_j)=T\sum_{\mathbf{p},\omega_n}G_{\mathbf{p}+\mathbf{q},\uparrow}^0(i\omega_n+i\nu_j)G^0_{-\mathbf{p},\downarrow}(-i\omega_n),
\eeq
where $G^0_{\mathbf{p},\sigma}(i\omega_n)=\left(i\omega_n-\xi_{\mathbf{p},\sigma}\right)^{-1}$
is the bare Green's function.
\par
Figures \ref{fig1} (a1) and (a2) show the temperature dependence of the ejection RF spectra $I_{\rm E}(\omega)$ at $(k_{\rm F}a)^{-1}=0$ and $(k_{\rm F}a)^{-1}=0.4$, respectively, and Figs.~ \ref{fig1} (b1) and (b2) show the reverse RF spectra $I_{\rm R}(\omega)$ in the same condition.
Their line-shapes are also reported in Figs. \ref{fig1} (c1) and (c2). 
In accordance with the different $\omega$-dependences, it follows that each RF spectrum shows different features associated with the properties of the system.
The ejection RF spectra reveal the occupied state at the thermal equilibrium
since impurities in the thermodynamic many-body state are directly transferred to 
the weakly-interacting final state.
At a low temperature, $I_{\rm E}(\omega)$ has a strong intensity near the attractive branch given by $-E_{\rm a}$ where $E_{\rm a}<0$ is the attractive polaron energy.
Such a peak is shifted towards the low-energy regime associated with a Boltzmann gas or repulsive branch with increasing the temperature.
We note that the peak energy slightly increases with increasing the temperature at $T\lesssim 0.8T_{\rm F}$ reflecting the increase of $|E_a|$. 
On the other hand, excitation properties of impurities are probed by
the reverse RF spectrum in which the weak-interacting initial state is transferred to the strongly-interacting minority state.  
In this regard, one can see the second peak near the repulsive branch associated with the repulsive polaron energy $E_{\rm r}>0$ in addition to the attractive branch around $\omega=E_{\rm a}$, as can be seen  from Fig.~\ref{fig1} (b2).
In contrast to the results at unitarity shown in Fig. \ref{fig1} (a1) and (b1) where the attractive polarons simply undergo the Boltzmann gas,
both RF spectra at $(k_{\rm F}a)^{-1}=0.4$ show the double peak structure and the effect of 
the repulsive branch remains up to $T/T_{\rm F}\sim 1$.
\par
\begin{figure}[t]
\begin{center}
\includegraphics[width=6cm]{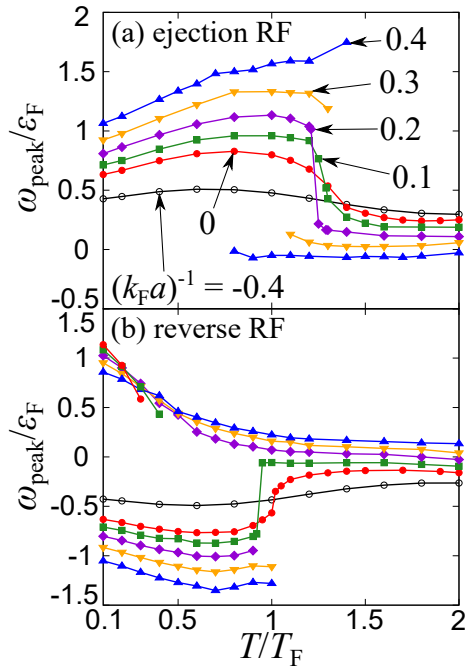}
\end{center}
\caption{Peak positions $\omega_{\rm peak}$ of (a) ejection and (b) reverse RF spectra at different interaction strengths. The common line symbols are used in each figure.
}
\label{fig2}
\end{figure}
To see how thermal evolution of the spectra is affected by the repulsive branch,
in Fig. \ref{fig2}, we plot the peak positions $\omega_{\rm peak}$ of (a) ejection and (b) reverse RF spectra at different impurity-bath interaction strengths. 
At the weak coupling $(k_{\rm F}a)^{-1}=-0.4$ where the repulsive branch is absent, regardless of the schemes of the RF spectroscopy,
one can see the smooth crossover from attractive polarons at the low-temperature regime  to the Boltzmann gas in the high-temperature regime.
By increasing the interaction, the crossover behavior becomes gradually sharper around $T/T_{\rm F}=1$ and changes to the transition-like jump at $(k_{\rm F}a)^{-1}\simeq 0.1$. 
In the ejection RF spectra, the single curve present  in the weak-coupling regime is split into the two curves at $(k_{\rm F}a)^{-1}\simeq 0.3$, and the temperature region where the double peaks coexist emerges.
On the other hand, we find different behaviors between the ejection and reverse RF spectroscopies due to the existence of the repulsive branch. 
In particular, the result for $I_{\rm E}(\omega)$ at $(k_{\rm F}a)^{-1}=0.4$ shows that
 the lower peak associated with the repulsive branch appears around $T/T_{\rm F}\simeq 0.8$ and the upper peak associated with the attractive branch disappears around $T/T_{\rm F}\simeq1.4$, since it is largely broadened and overlaps with the lower peak.
In contrast to the ejection RF, one can find the coexistence of two peaks in the low-temperature region of the reverse RF spectra. This is due to the fact that the peak of the repulsive branch evolves even at a low temperature as increasing the interaction.
In addition, as increasing the temperature, the repulsive peak corresponding to $E_{\rm r}$ decreases due to
the competition between the self-energy shift on the repulsive branch and the thermal agitation. 
We note that this behavior is in sharp contrast to the behavior of the repulsive polaron energy as a function of impurity concentration \cite{Tajima_NJP}.
Moreover, the peak of the attractive branch disappears around $T/T_{\rm F}=1$ beyond $(k_{\rm F}a)^{-1}\simeq 0.2$.
This indicates that the transition to the Boltzmann gas is deeply related to the existence of the repulsive polaron in the strong-coupling regime.
In the high-temperature limit, both RF spectra converge to the single peak.

\begin{figure}[t]
\begin{center}
\includegraphics[width=6cm]{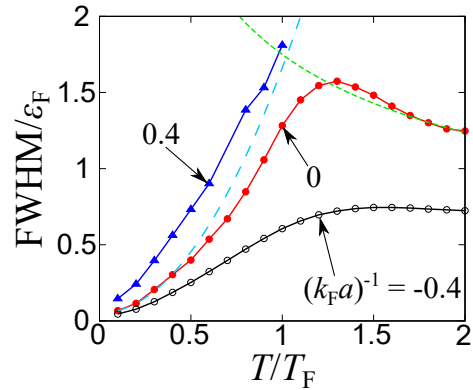}
\end{center}
\caption{FWHM of the ejection RF spectra at different interaction strengths. 
The dashed (long-dashed) line represents the high (low)-temperature fitting $\gamma/\varepsilon_{\rm F}=1.75\sqrt{T_{\rm F}/T}$ [$\gamma/\varepsilon_{\rm F}=1.61(T/T_{\rm F})^2$] at $(k_{\rm F}a)^{-1}=0$ \cite{Boiling,EnssZwerger,PRA92.053611}.
}
\label{fig3}
\end{figure}
\par
In Fig. \ref{fig3}, we show the full width at half maximum (FWHM) of ejection RF spectra, which is directly related to the decay rate of quasiparticles.
At the weak coupling $(k_{\rm F}a)^{-1}=-0.4$ and unitarity, FWHM has a maximum in the intermediate-temperature region.
In particular, the result at unitarity in the high-temperature regime is well fitted by the classical Boltzmann limit
$\gamma/\varepsilon_{\rm F}=1.75\sqrt{T_{\rm F}/T}$ \cite{Boiling,EnssZwerger,PRA92.053611}.
Thus, attractive polarons in fact undergo the Boltzmann gas regime beyond the maximum of FWHM.
While the repulsive polaron is expected to undergo such a classical regime at the strong coupling $(k_{\rm F}a)^{-1}=0.4$, FWHM is ill-defined because two broad peaks are overlapped each other [see Fig. \ref{fig1} (c2)].
On the other hand, we find the increase of FWHM in the low-temperature quantum regime as increasing the temperature at each interaction strength.
This growth gets stronger with increasing the interaction due to the collisional decoherence effects of attractive polarons 
at finite impurity concentration~\cite{PhysRevLett.118.083602}.
The quadratic behavior of the decay rate $\gamma/\varepsilon_{\rm F}=1.61(T/T_{\rm F})^2$ \cite{note1} is obtained at unitarity, which is consistent with the 
Fermi-liquid theory \cite{BruunCol}.
 \par
\begin{figure}[t]
\begin{center}
\includegraphics[width=6cm]{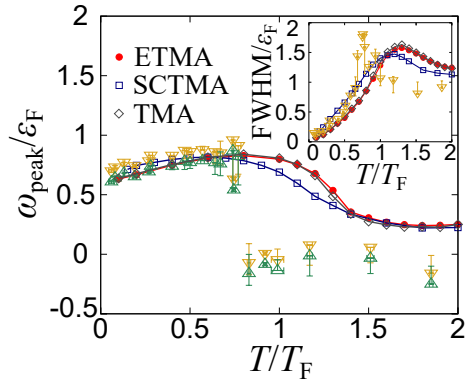}
\end{center}
\caption{Comparisons of peak positions and FWHM (inset) in ejection RF spectra between our results (filled circle: ETMA, open box: SCTMA, open diamond : TMA) and recent experiments at the unitarity limit.
The downward and upward triangles represent experimental value from Ref.~\cite{Boiling} and that with the subtraction of the mean-field shift with respect to the final state interaction given by $\Sigma_{\rm f}=0.09\varepsilon_{\rm F}$, respectively.
}
\label{fig4}
\end{figure}
We now compare our formalism with the recent experiment at unitarity~\cite{Boiling}.
To this end, let alone ETMA, 
we also perform two different diagrammatic approximations; the non-selfconsistent~\cite{Massignan2,Massignan3,Pietila}  and self-consistent $T$-matrix approximations~\cite{HaussmannSCTMA,PhysRevA.80.063612,FrankZwerger}, which are abbreviated as TMA and SCTMA, respectively. 
In Fig.~\ref{fig4}, we show the comparisons of peak positions and FWHM in the ejection RF spectra.
We note that while the TMA self-energy is obtained by replacing $G$ with $G_0$ in Eq. (\ref{eq3}), SCTMA is the approximation such that two $G^0$ in Eq. (\ref{eq5}) are replaced by $G$.
It is notable that our results for the spectral peak quantitatively reproduce the experiment (with the subtraction of the mean-field shift of final state interaction) up to 
$T/T_{\rm F}\simeq0.75$, where the sharp transition between attractive polarons and Boltzmann gas occurs.
While our results are consistent with the second-order virial expansion~\cite{Mulkerin}, we find the discrepancy between our results and the experiment around the transition temperature.  In addition,
the fitting coefficients of the decay rate in both high-temperature and low-temperature regimes deviates from the experimental values quantitatively.
In the low-temperature regime, the deviation of FWHM occurs in TMA and ETMA, which can be explained 
by a collisional decay process associated with dressed particles at finite impurity concentration as discussed in the context of the polaron-to-polaron decay process of the repulsive polaron \cite{PhysRevLett.118.083602,Massignan4}.
While TMA and ETMA do not include such a process, SCTMA does it through the dressed propagators in the pair susceptibility, Eq. (\ref{eq5}). 
Indeed, the quadratic fit of SCTMA gives $\gamma/\varepsilon_{\rm F}=2.95(T/T_{\rm F})^2$, which is close to the experimental result [$\gamma/\varepsilon_{\rm F}=2.71(T/T_{\rm F})^2$] \cite{Boiling}.
\par
\begin{figure}[t]
\begin{center}
\includegraphics[width=5.5cm]{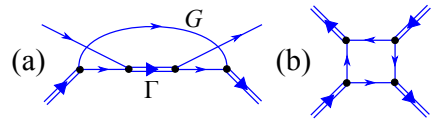}
\end{center}
\caption{Feynman diagrams describing examples of (a) three-body and (b) four-body correlations which are absent in TMA, where
the solid (double-solid) lines represent $G$ ($\Gamma$). 
While ETMA involves (a), SCTMA does both diagrams. }
\label{fig5}
\end{figure}

On the other hand, the discrepancy in the high-temperature regime is related to
multi-body correlations, which may be considered to be negligible in two-component Fermi gases due to the Pauli blocking effect. 
When the Fermi surface is thermally depleted, however, the Pauli-blocking effect is suppressed. Therefore,
beyond-two-body correlations can have a significant effect on the Fermi polarons at finite temperature as in the case of a Bose polaron \cite{MGHu,Jorgensen,Sun,Naidon,Yoshida,Blume}.
Since the experiment can be well reproduced by our theories in the low-temperature regime, where the Pauli-blocking effect emerges,
we argue that beyond-two-body correlations such as three-body scattering shift the critical interaction for the sharp transition towards the weaker coupling regime.
Furthermore, we uncover the difference among three different diagrammatic approaches in terms of three- and four-body correlations.
While TMA involves an atom-dimer scattering process within the Born level, 
ETMA and SCTMA  include the second-order three-body correlations as shown in Fig. \ref{fig5}(a).
In addition, SCTMA includes the lowest-order four-body correlations shown in Fig. \ref{fig5}(b), which are absent  both in TMA and ETMA. 
Indeed, SCTMA is closest to the experiment around $T=T_{\rm F}$ among these approaches.
Although our formalism does not fully explain the experiment, we find that the results are sensitive to the different approximations in the intermediate-temperature region, implying the importance of multi-body correlations.
This fact is consistent with the result of the virial expansion~\cite{Mulkerin,PRA92.053611}.
\par
In conclusion, we have investigated the thermal evolution of radio frequency spectra in an imbalanced Fermi gas system from the quantum degenerate state to the classical Boltzmann regime using the many-body $T$-matrix approximation.
We have predicted that the thermal evolution is classified into three types: smooth crossover, sharp transition, and double-peak coexistence, where the recent experiment in MIT has detected the second one at the unitarity limit.
These three types are transferred each other by shifting the impurity-bath interaction.
They can be detected in future experiments with finite scattering lengths.
We have  also revealed that the repulsive polarons play significant roles in the thermal evolution of the RF spectra.
From the quantitative viewpoint, our results perfectly reproduce the observed peak position of ejection RF spectra up to $T/T_{\rm F}\simeq 0.75$ but deviates from it in the high-temperature regime where the Pauli-blocking is absent. To explain the high-temperature regime, our results strongly suggest necessity of a selfconsistent formalism beyond 
the existing approximations, which contains multi-body correlations.
\par
We thank Z. Yan and M. W. Zwierlein for providing us with their experimental data for comparison.
H. T. thanks F. Scazza, M. Zaccanti, D. Kagamihara, and R. Sato for useful discussions.
H. T. is supported by a Grant-in-Aid for JSPS fellows (No. 17J03975).
S. U. is supported by JSPS KAKENHI Grant Number JP17K14366 and 
by a Waseda University Grant for Special Research Projects (No. 2018S-209).
This work was partially supported by RIKEN iTHEMS program.

\bibliographystyle{apsrev4-1}

\begin{thebibliography}{99}
\bibitem{PRD60.091503} Y. Nakahara, M. Asakawa, and T. Hatsuda, Phys. Rev. D \textbf{60}, 091503 (1999).
\bibitem{PPNP46.459} M. Asakawa, T. Hatsuda, and Y. Nakahara, Prog. Part. Nucl. Phys. \textbf{46}, 459 (2001).
\bibitem{RMP82.1095} E. Klempt and J.-M. Richard, Rev. Mod. Phys. \textbf{82}, 1095 (2010).

\bibitem{GW170817} B. P. Abbott {\it et al}., Phys. Rev. Lett. \textbf{119}, 161101 (2017).
\bibitem{AbbottAstrophys} B. P. Abbott {\it et al}., Astrophys. J. \textbf{848}, L12 (2017).

\bibitem{PRL80.149} C. Renner, B. Revaz, J.-Y. Genoud, K. Kadowaki, and \O. Fischer, Phys. Rev. Lett. \textbf{80}, 149 (1998).
\bibitem{RMP75.473} A. Damascelli, Z. Hussain, and Z.-X. Shen, Rev. Mod. Phys. \textbf{75}, 473 (2003).
\bibitem{RMP79.353} \O. Fischer, M. Kugler, I. Maggio-Aprile, C. Berthod, and C. Renner, Rev. Mod. Phys. \textbf{79}, 353 (2007).
\bibitem{Lu2012} D. Lu, I. N. Vishik, M. Yi, Y. Chen, R. G. Moore, and Z.-X. Shen, Annu. Rev. Cond. Mat. Phys. \textbf{3}, 129 (2012).
\bibitem{torma} P. T\"{o}rm\"{a}, "Spectroscopies - theory", in {\it Quantum Gas Experiments - Exploring Many-Body States},
edited by P. T\"{o}rm\"{a} and K. Sengstock (Imperial Colledge Press, London, 2015).


\bibitem{steinhauer} J. Steinhauer, R. Ozeri, N. Katz, and N. Davidson, Phys. Rev. Lett. \textbf{88}, 120407 (2002).
\bibitem{Hoinka} S. Hoinka, P. Dyke, M. G. Lingham, J. J. Kinnunen, G. M. Bruun, and C. J. Vale, Nat. Phys. \textbf{13}, 943 (2017).
\bibitem{soferle} T. Stoferle, H. Moritz, C. Schori, M. Kohl, and T. Esslinger, Phys. Rev. Lett. \textbf{92}, 130403 (2004).

\bibitem{Chin} C. Chin, M. Bartenstein, A. Altmeyer, S. Riedl, S. Jochim, J. Hecker Denschlag, and R. Grimm, Science \textbf{305}, 1128 (2004).
\bibitem{Stewart} J. T. Stewart, J. P. Gaebler, and D. S. Jin, Nature \textbf{454}, 744 (2008).
\bibitem{Gaebler} J. P. Gaebler, J. T. Stewart, T. E. Drake, D. S. Jin, A. Perali, P. Pieri, and G. C. Strinati, Nat. Phys. \textbf{6}, 569 (2010).
\bibitem{Sagi} Y. Sagi, T. E. Drake, R. Paudel, R. Chapurin, and D. S. Jin Phys. Rev. Lett. \textbf{114}, 075301 (2015). 
\bibitem{Schirotzek} A. Schirotzek, Y. Shin, C. H. Schunk, and W. Ketterle, Phys. Rev. Lett. \textbf{101}, 140403 (2008).
\bibitem{Behrle} A. Behrle, T. Harrison, J. Kombe, K. Gao, M. Link, J.-S. Bernier, C. Kollath, and M. K\"{o}l, Nat. Phys. \textbf{14}, 781 (2018).

\bibitem{Lompe} T. Lompe, T. B. Ottenstein, F. Serwane, A. N. Wenz, G. Z\"{u}rn, and S. Jochim, Science \textbf{330}, 940 (2010).
\bibitem{Nakajima} S. Nakajima, M. Horikoshi, T. Mukaiyama, P. Naidon, and M. Ueda, Phys. Rev. Lett. \textbf{106}, 143201 (2011).



\bibitem{Schirotzek0902} A. Schirotzek, C.-H. Wu, A. Sommer, and M. W. Zwierlein,
Phys. Rev. Lett. \textbf{102}, 230402 (2009).
\bibitem{nascimbene} S. Nascimb\`ene, N. Navon, K. J.  Jiang,  L. Tarruell, M. Teichmann, J. McKeever, F. Chevy,  and C. Salomon,
Phys. Rev. Lett. \textbf{103}, 170402 (2009).
\bibitem{Kohstall1112} C. Kohstall, M. Zaccanti, M. Jag, A. Trenkwalder,
P. Massignan, G. M. Bruun, F. Schreck, and R. Grimm,
Nature \textbf{485}, 615 (2011).
\bibitem{koschorreck2012} M. Koschorreck, D. Pertot, E. Vogt, B. Fr\"{o}hlich, M. Feld,
and M. K\"{o}hl, Nature \textbf{485}, 619 (2012).
\bibitem{Cetina96} M. Cetina, M. Jag, R. S. Lous, I. Fritsche, J. T. M. Walraven,
R. Grimm, J. Levinsen, M. M. Parish, R. Schmidt,
M. Knap, and E. Demler, Science \textbf{354}, 96 (2016).
\bibitem{PhysRevLett.118.083602} F. Scazza, G. Valtolina, P. Massignan, A. Recati, A. Amico,
A. Burchianti, C. Fort, M. Inguscio, M. Zaccanti,
and G. Roati, Phys. Rev. Lett. \textbf{118}, 083602 (2017).


\bibitem{massignan} P. Massignan, M. Zaccani, G. M. Bruum, Rep. Prog. Phys. \textbf{77},
034401 (2014).
\bibitem{Combescot2008}R. Combescot and S. Giraud, Phys. Rev. Lett. \textbf{101},
050404 (2008).
\bibitem{Bruum2010} G. M. Bruun and P. Massignan, Phys. Rev. Lett. \textbf{105},
020403 (2010).
\bibitem{PhysRevA.85.021602} R. Schmidt, T. Enss, V. Pietil\"{a}, and E. Demler, Phys.
Rev. A \textbf{85}, 021602 (2012).
\bibitem{PhysRevA.85.033631} J. E. Baarsma, J. Armaitis, R. A. Duine, and H. T. C.
Stoof, Phys. Rev. A 85, 033631 (2012).
\bibitem{PhysRevA.74.063628} F. Chevy, Phys. Rev. A \textbf{74}, 063628 (2006).
\bibitem{Combescot2007} R. Combescot, A. Recati, C. Lobo, and F. Chevy, Phys.
Rev. Lett. \textbf{98}, 180402 (2007).
\bibitem{Punk2009} M. Punk, P. T. Dumitrescu, and W. Zwerger, Phys. Rev.
A \textbf{80}, 053605 (2009).
\bibitem{PhysRevA.80.033607} C. Mora and F. Chevy, Phys. Rev. A \textbf{80}, 033607 (2009).
\bibitem{PhysRevA.81.041602} X. Cui and H. Zhai, Phys. Rev. A \textbf{81}, 041602 (2010).
\bibitem{PhysRevLett.106.166404} C. J. M. Mathy, M. M. Parish, and D. A. Huse, Phys.
Rev. Lett. \textbf{106}, 166404 (2011).
\bibitem{Trefzger2012} C. Trefzger and Y. Castin, Phys. Rev. A \textbf{85}, 053612
(2012).
\bibitem{Prokofev20081} N. Prokof'ev and B. Svistunov, Phys. Rev. B \textbf{77}, 020408
(2008).
\bibitem{Prokofev20082} N. V. Prokof'ev and B. V. Svistunov, Phys. Rev. B \textbf{77},
125101 (2008).
\bibitem{Vlietinck2013} J. Vlietinck, J. Ryckebusch, and K. Van Houcke, Phys.
Rev. B \textbf{87}, 115133 (2013).
\bibitem{Kroiss2015} P. Kroiss and L. Pollet, Phys. Rev. B \textbf{91}, 144507 (2015).
\bibitem{PhysRevA.94.051605} O. Goulko, A. S. Mishchenko, N. Prokof'ev, and B. Svistunov,
Phys. Rev. A \textbf{94}, 051605 (2016).

\bibitem{Schmidt:2011zu} R. Schmidt and T. Enss, Phys. Rev. A \textbf{83}, 063620 (2011).
\bibitem{PhysRevA.95.013612} K. Kamikado, T. Kanazawa, and S. Uchino, Phys. Rev.
A \textbf{95}, 013612 (2017).
\bibitem{Bulgac} A. Bulgac, J. E. Drut, and P. Magierski, Phys. Rev. A \textbf{78}, 023625 (2008).
\bibitem{Goulko2} O. Goulko and M. Wingate, Phys. Rev. A \textbf{82}, 053621 (2010).

\bibitem{boxpot} B. Mukherjee, Z. Yan, P. B. Patel, Z. Hadzibabic, T. Yefsah, J. Struck, and M. W. Zwierlein, Phys. Rev. Lett. \textbf{118}, 123401 (2017).
\bibitem{Boiling} Z. Yan, P. B. Patel, B. Mukherjee, R. J. Fletcher, J. Struck, and M. W. Zwierlein, arXiv:1811.00481v1 [cond-mat.quant-gas].
\bibitem{Massignan2} P. Massignan, G. M. Bruun, and H. T. C. Stoof, Phys. Rev. A \textbf{77}, 031601(R) (2008).
\bibitem{Massignan3} P. Massignan, G. M. Bruun, and H. T. C. Stoof, Phys. Rev. A \textbf{78}, 031602(R) (2008).
\bibitem{Pietila} V. Pietil\"{a}, Phys. Rev. A \textbf{86}, 023608 (2012).
\bibitem{Doggen} E. V. H. Doggen and J. J. Kinnunen, Phys. Rev. Lett. \textbf{111}, 025302 (2013). 
\bibitem{Hu} H. Hu, B. C. Mulkerin, J. Wang, and X.-J. Liu, Phys. Rev. A \textbf{98}, 013626 (2018).
\bibitem{Tajima_NJP} H. Tajima and S. Uchino, New J. Phys. \textbf{20}, 073048 (2018).
\bibitem{Edward} W. E. Liu, J. Levinsen, and M. M. Parish, arXiv:1805.10013 [cond-mat.quant-gas].
\bibitem{Mistakidis} S. I. Mistakidis, G. C. Katsimiga, G. M. Koutentakis, and P. Schmelcher, arXiv:1808.00040v2 [cond-mat.quant-gas].
\bibitem{Mulkerin} B. C. Mulkerin, X.-J. Liu, and H. Hu, arXiv:1808.07671v1 [cond-mat.quant-gas].

\bibitem{noteC} To obtain a precise contact parameter from the high-frequency line shape as done in the experiment \cite{Boiling},
we require the Matsubara Green's function up to the very large number of the Matsubara frequency in our theoretical framework.

\bibitem{PhysRevA.86.043622}T. Kashimura, R. Watanabe, and Y. Ohashi, Phys. Rev.
A \textbf{86}, 043622 (2012).
\bibitem{PhysRevA.89.033617} H. Tajima, T. Kashimura, R. Hanai, R. Watanabe, and
Y. Ohashi, Phys. Rev. A \textbf{89}, 033617 (2014).
\bibitem{PhysRevA.93.013610} H. Tajima, R. Hanai, and Y. Ohashi, Phys. Rev. A \textbf{93},
013610 (2016).
\bibitem{PhysRevA.96.033614} H. Tajima, R. Hanai, and Y. Ohashi, Phys. Rev. A \textbf{96},
033614 (2017).
\bibitem{PhysRevA.95.043625} H. Tajima, P. van Wyk, R. Hanai, D. Kagamihara, D. Inotani,
M. Horikoshi, and Y. Ohashi, Phys. Rev. A \textbf{95},
043625 (2017).
\bibitem{PhysRevX.7.041004}M. Horikoshi, M. Koashi, H. Tajima, Y. Ohashi, and
M. Kuwata-Gonokami, Phys. Rev. X \textbf{7}, 041004 (2017).

\bibitem{EnssZwerger} T. Enss, R. Haussman, and W. Zwerger, Ann. Phys. (N. Y.). \textbf{326}, 770 (2011).

\bibitem{note1} There is an offset $\simeq0.05$ associated with $\delta$ used for the analytic continuation as well as the finite-impurity-density effect.

\bibitem{BruunCol} G. M. Bruun, A. Recati, C. J. Pethick, H. Smith, and S. Stringari, Phys. Rev. Lett. \textbf{100}, 240406 (2008).



\bibitem{HaussmannSCTMA} R. Haussmann, W. Rantner, S. Cerrito, and W. Zwerger, Phys. Rev. A \textbf{75}, 023610 (2007).
\bibitem{PhysRevA.80.063612} R. Haussmann, M. Punk, and W. Zwerger, Phys. Rev.
A \textbf{80}, 063612 (2009).
\bibitem{FrankZwerger} B. Frank, J. Lang, and W. Zwerger, arXiv:1804.03035 [cond-mat.quant-gas].

\bibitem{Massignan4} P. Massignan and G. M. Bruun, Eur. Phys. J. D \textbf{65}, 83 (2011).

\bibitem{MGHu} M.-G. Hu, M. J. Van de Graaff, D. Kedar, J. P. Corson, E. A. Cornell, and D. S. Jin, Phys. Rev. Lett. \textbf{117}, 055301 (2016).
\bibitem{Jorgensen} N. B. J\o rgensen, L. Wacker, K. T. Skalmstang, M. M. Parish, J. Levinsen, R. S. Christensen, G. M. Bruun, and J. J. Arlt, Phys. Rev. Lett. \textbf{117}, 055302 (2016).

\bibitem{Sun} M. Sun, H. Zhai, and X. Cui, Phys. Rev. Lett. \textbf{119}, 013401 (2017).
\bibitem{Naidon} P. Naidon, J. Phys. Soc. Jpn. \textbf{87}, 043002 (2018).
\bibitem{Yoshida} S. M. Yoshida, S. Endo, J. Levinsen, and M. M. Parish, Phys. Rev. X \textbf{8}, 011024 (2018).
\bibitem{Blume} D. Blume, Phys. Rev. A \textbf{99}, 013613 (2019).

\bibitem{PRA92.053611} M. Sun and X. Leyronas, Phys. Rev. A \textbf{92}, 053611 (2015).


\end{thebibliography}

\end{document}